\documentclass[pra,superscriptaddress,longbibliography,twocolumn]{revtex4-1}
\usepackage[T1]{fontenc}
\usepackage{amsmath}
\usepackage{amssymb}
\usepackage{mathtools}
\usepackage[normalem]{ulem}
\usepackage{graphicx}
\usepackage{datetime}
\usepackage{color}
\usepackage{newtxtext,newtxmath}
\usepackage{microtype}
\usepackage{braket}
\usepackage{xr}
\usepackage{natbib}
\usepackage{hyperref}
\hypersetup{%
    bookmarksopen=false,
    bookmarksnumbered=true,
    pdfnewwindow=true,
    unicode=false,
    colorlinks=true,%
    citecolor=blue,
    linkcolor=black,
    urlcolor=blue,
    filecolor=blue
    }  
\newcommand{\aDq}[1]{\big\langle\tilde{D}_{#1}\big\rangle}
\newcommand{\ar}{\left<r\right>} 

\newcommand{\Dq}[1]{\tilde{D}_{#1}}
\newcommand{\eps}{\varepsilon}

\newcommand{\mrm}[1]{\mathrm{#1}}
\newcommand{\NN}{\mathcal{N}}

\renewcommand{\P}{\mathcal{P}}
\newcommand{\R}{\mathbb{R}}

\DeclareMathOperator{\var}{var}
\newcommand{\vDq}[1]{\var\big(\tilde{D}_{#1}\big)}
\renewcommand{\vec}[1]{\boldsymbol{#1}}
\begin{document}
\newcommand{\figdir}{.}
\newcommand{\freiburg}{Physikalisches Institut, Albert-Ludwigs-Universit\"{a}t Freiburg, Hermann-Herder-Stra{\ss}e 3, D-79104, Freiburg, Germany}
\newcommand{\usal}{Departamento de F\'isica Fundamental, Universidad de Salamanca, E-37008 Salamanca, Spain}
\newcommand{\eucor}{EUCOR Centre for Quantum Science and Quantum Computing, Albert-Ludwigs-Universit\"{a}t Freiburg, Hermann-Herder-Stra{\ss}e 3, D-79104, Freiburg, Germany}

\title{Optimal route to quantum chaos in the Bose-Hubbard model}
\author{Lukas Pausch}
\affiliation{\freiburg}
\author{Andreas Buchleitner}
\email[]{a.buchleitner@physik.uni-freiburg.de}
\affiliation{\freiburg}
\affiliation{\eucor}
\author{Edoardo G.\ Carnio}
\affiliation{\freiburg}
\affiliation{\eucor}
\author{Alberto Rodr\'iguez}
\email[]{argon@usal.es}
\affiliation{\usal}

\begin{abstract}
  The dependence of the chaotic phase of the Bose-Hubbard Hamiltonian on particle number $N$, system size $L$ and particle density is investigated in terms of spectral and eigenstate features. We analyze the development of the chaotic phase as the limit of infinite Hilbert space dimension is approached along different directions, and show that the fastest route to chaos is the path at fixed density $n\lesssim1$. The limit $N\to\infty$ at constant $L$ leads to a slower convergence of the chaotic phase towards the random matrix theory benchmarks. In this case, from the distribution of the eigenstate generalized fractal dimensions, the chaotic phase becomes more distinguishable from random matrix theory for larger $N$, in a similar way as along trajectories at fixed density.
\end{abstract}
\maketitle
\section{Introduction}

Quantum chaos \cite{Haake2004} plays a central role in 
present investigations of complex many-particle dynamics, with systems of trapped ultracold atoms as prominent experimental examples 
\cite{Ronzheimer2013a,meinert2014,Preiss2015,Islam2015,Kondov2015a,Schreiber2015b,Choi2016a,Bordia2016,Meinert2016a,Kaufman2016,Bordia2017a,Rispoli2018a,Kohlert2019}. 
Many-particle chaotic dynamical behaviour entails, e.g., fast entanglement generation and quantum information spreading, whose control is a requirement in current experimental quantum simulation platforms \cite{Garttner2017,Joshi2020,Berke2020,Mi2021}.
  
In the quantum realm, the potential emergence of chaos in a certain parametric regime of the system's Hamiltonian is ascertained from the comparison of its spectral or eigenvector features against well-known benchmarks dictated by random matrix theory (RMT) \cite{Casati1980,bgs84,Bohigas1984a,Berry1985,Giannoni89,Izrailev1990,Guhr1998,Muller2004,Borgonovi2016}. Furthermore, in a many-particle system, the existence of the chaotic regime is assessed by checking its persistence as the particle number $N$ is increased, i.e., in the limit $N\to\infty$, which also encompasses an infinite dimensional Hilbert space and can be interpreted as a many-particle semiclassical limit (for large particle density), $\hbar_\textrm{eff}=N^{-1}\to 0$ (see, for instance, Ref.~\cite{Richter2022}).
 
Interacting bosons in a lattice, which are conveniently  described by the Bose-Hubbard Hamiltonian \cite{Lewenstein2007,Bloch2008,Cazalilla2011,Krutitsky2016}, constitute an archetypical system to study the appearance of quantum chaos 
\cite{Kolovsky2004,Kollath2010,Beugeling2014,Beugeling2015,Beugeling2015c,Dubertrand2016,Fischer2016a,Beugeling2018} and its dynamical consequences \cite{Buchleitner2003,Kollath2007,Venzl2009,Roux2009,Roux2010,Biroli2010b,Sorg2014,Dufour2020} that is also amenable to treatment with modern semiclassical techniques in the above mentioned limit 
\cite{Hiller2006,Hiller2009,Engl2014,Engl2015,Tomsovic2018,Rammensee2018}.
In Refs.~\cite{Pausch2020,Pausch2021}, we presented a detailed characterization of the chaotic phase of the Bose-Hubbard model at unit particle density, and demonstrated the ability of RMT to capture coarse-grained features of the eigenvector structure in Fock space, as well as the possibility to discriminate the model's chaotic phase from that of random two-body Hamiltonians. 
Nonetheless, to the best of our knowledge, no systematic study of the development of quantum chaos as the limit of infinite Hilbert space dimension is approached along different directions has been carried out. 

Here, we present a thorough analysis of the dependence of the Bose-Hubbard Hamiltonian's chaotic phase on particle number, system size, and density. 
In particular, we investigate whether there is a quantifiable difference between the limits $N\to\infty$ at fixed system size and along trajectories at fixed bosonic density in relation to RMT benchmarks, and whether an optimal route to quantum chaos exists. 

The manuscript is organized as follows. In Sec.~\ref{sec:chaoticphase}, we give the particulars of the model and of the tools used to identify the chaotic phase. The influence of particle number, system size and 
density 
on the chaotic regime is scrutinized in Sec.~\ref{sec:chaosnu}, where we demonstrate the existence of an optimal route towards quantum chaos. The difference between the convergence to eigenstate ergodicity along the limits $N\to\infty$ at fixed system size and at constant density is discussed in Sec.~\ref{sec:pdfs}, followed by our final remarks in Sec.~\ref{sec:conclusions}.

\section{Characterization of the chaotic phase of the Bose-Hubbard Hamiltonian}
\label{sec:chaoticphase}
The one-dimensional Bose-Hubbard Hamiltonian (BHH) reads \cite{Lewenstein2007,Bloch2008,Cazalilla2011,Krutitsky2016}
\begin{equation}
   H = -J\sum_{j=1}^L\left(a^\dagger_j a_{j+1} + a^\dagger_{j+1} a_j\right) + \frac{U}{2} \sum_{j=1}^L a_j^\dagger a_j^\dagger a_j a_j,
   \label{eq:BHH}
\end{equation}
in terms of the standard operators $a_j^{(\dagger)}$ associated with Wannier orbitals localized at each lattice site. The bosonic system is then specified by the total particle number $N$, the number $L$ of spatial modes, the nearest-neighbour tunneling energy $J$, and the repulsive two-particle interaction energy $U>0$.

Besides time reversal invariance and particle number conservation, the existence of additional symmetries is determined by the boundary conditions of the lattice. Here, for simplicity, we only consider hard-wall boundary conditions, corresponding to setting $a^{(\dagger)}_{L+1}:= 0$ in Eq.~\eqref{eq:BHH}. In this case, $H$ exhibits reflection symmetry about the lattice centre ($[H,\Pi]=0$ where $\Pi$ denotes the reflection operation) and Hilbert space can be written as the direct sum 
\begin{equation}
\mathcal{H}=\mathcal{H}^+\oplus \mathcal{H}^-,
\end{equation}
with $\mathcal{H}^\pm$ the symmetric (even parity) and antisymmetric (odd parity) subspace, respectively.

The BHH is integrable in the limit of vanishing tunneling ($J=0$) as well as in the many-particle non-interacting limit, corresponding to $U=0$. At those limits, one can find as many independent and commuting conserved observables as underlying degrees of freedom, which are fixed by the number $L$ of spatial modes \cite{Pausch2022}. Integrability hence implies that the eigenstates of $H$ are Fock states uniquely identified by $L$ quantum numbers. For $J=0$, for instance, the eigenstates of $\mathcal{H}^-$, 
\begin{equation}
\ket{\vec{n}} = \frac{1}{\sqrt{2}}\left(1-\Pi\right)\ket{n_1,\ldots,n_L}, 
\label{eq:intbasis}
\end{equation}
are characterized by the eigenvalues of the $L$ onsite number operators $n_j = a_j^\dagger a_j$. 
For non-vanishing $J$ and $U$, the BHH is non-integrable and exhibits a chaotic phase visible in the spectral and eigenvector properties \cite{Kolovsky2004,Kollath2010,Dubertrand2016,Pausch2020,Pausch2021,Pausch2022}. 

The system's spectral and eigenvector features for varying $N$ and $L$ can be conveniently compared in terms of the \emph{scaled (with respect to the system's total spectral width) energy}
\begin{equation}
\eps = (E-E_\mrm{min})/(E_\mrm{max}-E_\mrm{min}),
\label{eq:Edef}
\end{equation}
where $E_\mrm{min}$, $E_\mrm{max}$ denote respectively the lowest and highest eigenenergies of $H$, 
and the \emph{scaled tunneling strength},
\begin{equation}
  \eta= J/UN.
  \label{eq:etadef}
\end{equation}
As we demonstrated in Refs.~\cite{Pausch2020,Pausch2021,Pausch2022}, the emergence of the chaotic phase at fixed scaled energy $\eps$ is dictated by $\eta$.
This can be understood from the behaviour of the spectrum boundaries. While the upper bound of the tunneling term in the Hamiltonian scales as $JN$, i.e., linearly with particle number, the maximal interaction energy reads $UN(N-1)/2$. As a function of $J/U$, one therefore expects a change from $(E_\mrm{max}-E_\mrm{min})\sim N^2$ to $(E_\mrm{max}-E_\mrm{min})\sim N$ for sufficiently high $J/U$. From the comparison of the spectral widths of the tunneling and interacting terms in Eq.~\eqref{eq:BHH}, the centre of such crossover may be estimated as $\eta_*\approx (1-\min(N,L)^{-1})/8$. 
For $\eta\gg\eta_*$, the scaled energy is thus equivalent to the
energy per particle.
For other $\eta$ values, 
however, approaching the thermodynamic limit at constant $\eps$ requires states whose energy scales quadratically with the particle number, i.e.,~whose interaction energy does so.  
For fixed $J/U$ such system configurations would end up being dominated by the interaction energy and thus in the non-ergodic phase of the system. A reduction of the interaction strength by $N$ is then necessary to ensure that both terms in the Hamiltonian scale linearly with particle number, and hence to observe the transition into the chaotic phase. Such rescaling of $U$ ensues, for instance, in the semiclassical limit, i.e., for increasing particle density, where the two parameters that govern the system dynamics are $E/UN^2$ and $\eta$ \cite{Hiller2006,Hiller2009,Dubertrand2016}. 
Notwithstanding, for system configurations whose interaction energy scales linearly with $N$ as the thermodynamic limit is approached,
and hence the energy per particle is fixed, 
the onset of the chaotic phase should be determined by the bare ratio $J/U$.

The characterization of spectral chaos in the sense of random matrix theory is based on the analysis of short-range features of the energy spectrum \cite{Haake2004,bgs84,Berry1985,Giannoni89,Guhr1998,Muller2004}, which is most conveniently carried out in terms of the 
level spacing ratios $r_n$ \cite{Oganesyan2007, Pal2010,Atas2013c}, 
\begin{align}
  r_n = \min\left(\frac{s_{n+1}}{s_n}, \frac{s_n}{s_{n+1}}\right) \in[0,1],
  \label{eq:rn}
\end{align}
where $s_n = E_{n+1}-E_n$ is the $n$th level spacing.
For the Gaussian orthogonal random-matrix ensemble (GOE), the analytic approximation to the distribution of $r$ reads \cite{Atas2013c}
\begin{equation}
  P_\mrm{GOE}(r)=\frac{27}{4}\frac{r+r^2}{(1+r+r^2)^{5/2}},
  \label{eq:prgoe}
\end{equation}
which yields the mean level spacing ratio $\ar_{\mrm{GOE}} = 4-2\sqrt{3}\approx 0.536$, in good agreement with $\ar_\mathrm{GOE}\approx 0.5307$ obtained from large-scale numerics \cite{Atas2013c}.

\begin{figure}
  \includegraphics[width=.9\columnwidth]{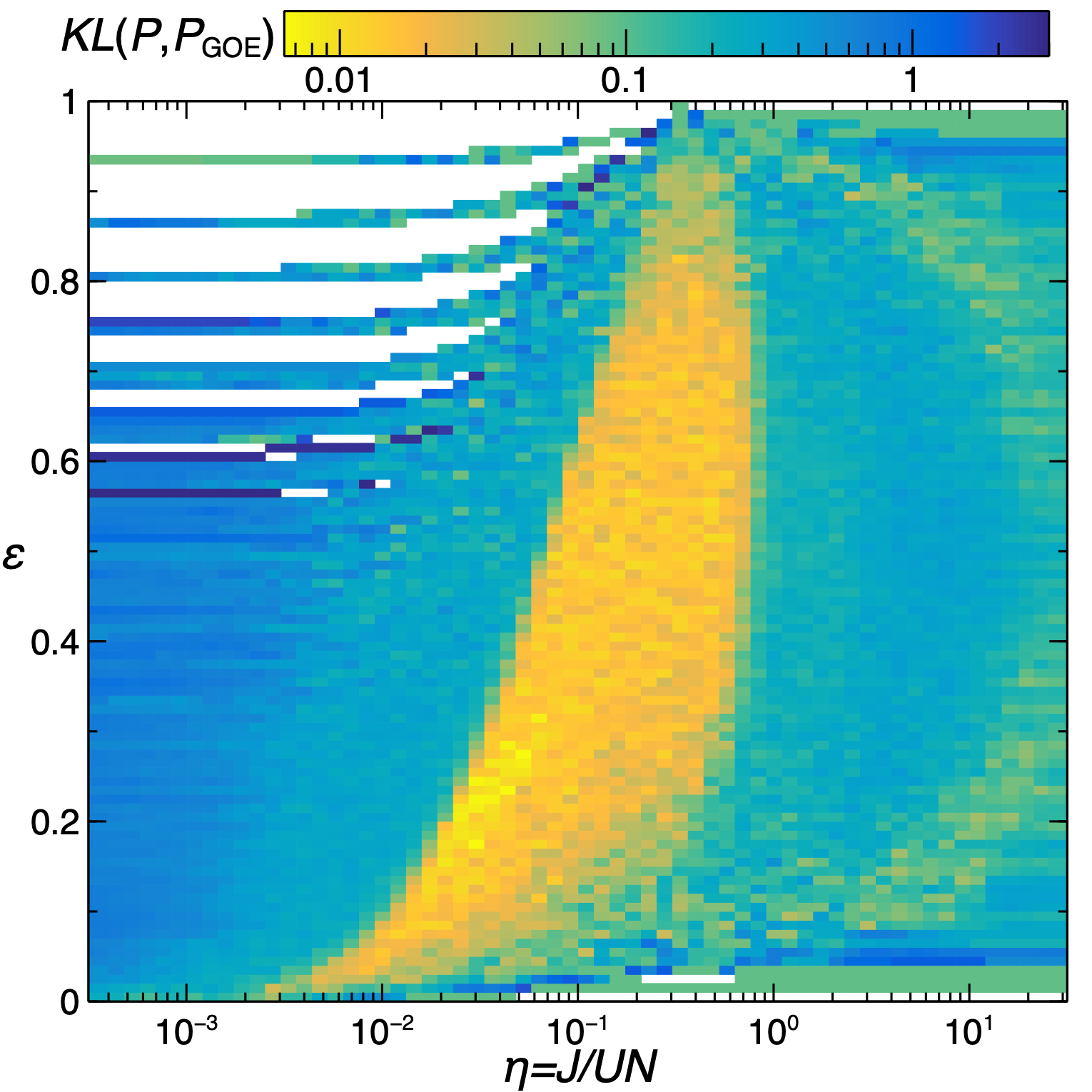}
  \caption{Perspective of the chaotic phase in the $(\eps,\eta)$ plane for the subspace $\mathcal{H}^-$ of the BHH for $L=5$ and $N=36$ (Hilbert space size $\mathcal{N}=45\,600$), as revealed by the distance between the distribution $P(r)$ of the $r$ statistic [Eq.~\eqref{eq:rn}] and the GOE distribution [Eq.~\eqref{eq:prgoe}], measured by the Kullback-Leibler divergence [Eq.~\eqref{eq:KL}]. Compare against the upper left panel of Fig.~\ref{fig:DPfixedHdim}.}
  \label{fig:prdiff}
\end{figure}

In Fig.~\ref{fig:prdiff}, we unveil the spectrally chaotic phase of the BHH as a function of $\eps$ and $\eta$ from the comparison of the $P(r)$ distribution against Eq.~\eqref{eq:prgoe}, quantified by the Kullback-Leibler divergence \cite{KullbackLeibler,InformationTheory}, 
\begin{align}
	KL\left(P,P_\textrm{GOE}\right) = \int_{0}^{1} P(r) \ln \left(\frac{P(r)}{P_\textrm{GOE}(r)}\right) \; \mathrm{d}r.
	\label{eq:KL}
\end{align}
For this analysis the energy axis is homogeneously discretized in 100 intervals, and for each $\eta$ the $P(r)$ distribution is constructed from the energy levels lying within each interval. The results highlight a tilted and elongated region along the energy axis within which $P(r)$ shows agreement with random matrix theory, in accord with the description obtained from the comparison of the average value $\langle r \rangle$ alone [cf.~Fig.~\ref{fig:DPfixedHdim}]. 
(An example of the evolution of $\ar$ versus $\eta$ after averaging over the inner part of the spectrum is shown in Fig.~3 of Ref.~\cite{Pausch2021}.)
The characteristic flow of the chaotic region's left boundary from low $\eps$ and low $\eta$ towards higher scaled energy and tunneling strength can be traced back to the trajectory of the maximum of the density of states in this parameter space \cite{Pausch2021,Pausch2022}.  

The formation of the chaotic phase also bears a specific and notable change in the eigenstates: Spectral chaos unambiguously correlates with the emergence of \emph{extended ergodic states} in Hilbert space in the thermodynamic limit, as we showed in Refs.~\cite{Pausch2020,Pausch2021}. A basic characterization of the eigenstate structure in a chosen basis is provided by the generalized fractal dimensions (GFDs) $D_q\in[0,1]$ \cite{Halsey1986,Nakayama2003,Rodriguez2010,Rodriguez2011}, which determine the scaling of the moments of the eigenstate's distribution of intensities as the Hilbert space size $\mathcal{N}$ becomes asymptotically large: $\sum_{\alpha}|\psi_\alpha|^{2q}\sim \mathcal{N}^{-(q-1)D_q}$ for $q\in\R^+$, where $\psi_\alpha$ are the eigenstate amplitudes in the basis $\{\ket{\alpha}\}$ 
[which here will be given by the states in Eq.~\eqref{eq:intbasis}].
Whereas vanishing GFDs indicate localization, and non-zero $q$-dependent GFDs signal a multifractal structure (which seems to be ubiquitous in many-body Hilbert spaces \cite{Atas2012,Atas2014,Luitz2014,Luitz2015,Torres-Herrera2017,Serbyn2017,Lindinger2019,Backer2019,Mace2019,Luitz2020,Pietracaprina2021}), an extended ergodic eigenstate, i.e., equipartition of the state over all basis elements as $\mathcal{N}\to\infty$, has associated GFDs $D_q=1$ for all $q$. 

We define the finite-size generalized fractal dimensions as 
\begin{align}
  \Dq{q} = -\frac{1}{q-1} \frac{\ln \sum_\alpha |\psi_\alpha|^{2q}}{\ln \NN},
  \label{eq:GFD}
\end{align}
whose asymptotic limits provide the GFDs, $D_q = \lim_{\NN\to \infty} \Dq{q}$.
We specifically consider 
\begin{align}
\Dq{1}&=\lim_{q\to 1}\Dq{q}=- \sum_{\alpha}\left|\psi_{\alpha}\right|^2\ln\left|\psi_{\alpha}\right|^2/\ln\NN, 
\end{align}
which is determined by the information entropy of the eigenstate. 

The emergence of the chaotic phase correlates with the occurrence of high $\Dq{q}$ values that approach the ergodic limit for increasing $\mathcal{N}$. Most interestingly, the fluctuation of the finite-size GFDs among near-in-energy eigenstates, characterized by $\vDq{q}$, is an extremely sensitive eigenstate-based probe of quantum chaos (cf.~Fig.~\ref{fig:DPfixedHdim}) that further shows a basis independent qualitative behaviour  \cite{Pausch2020,Pausch2021,Pausch2022}. 

\section{Dependence of the chaotic phase on $\boldsymbol{N}$, $\boldsymbol{L}$, and filling factor}
\label{sec:chaosnu}
To analyze how the chaotic phase evolves as a function of the number $N$ of bosons and $L$ of spatial modes, we obtain full spectra of the BHH numerically using exact diagonalization, and also calculate eigenstates and eigenenergies around chosen target energies \cite{Pietracaprina2018,petsc-user-ref,slepc}. We restrict ourselves to the Hilbert subspace $\mathcal{H}^-$ of the BHH with hard-wall boundary conditions, and work in the basis spanned by the states in Eq.~\eqref{eq:intbasis}. 

\begin{figure}
 \includegraphics[width=.95\columnwidth]{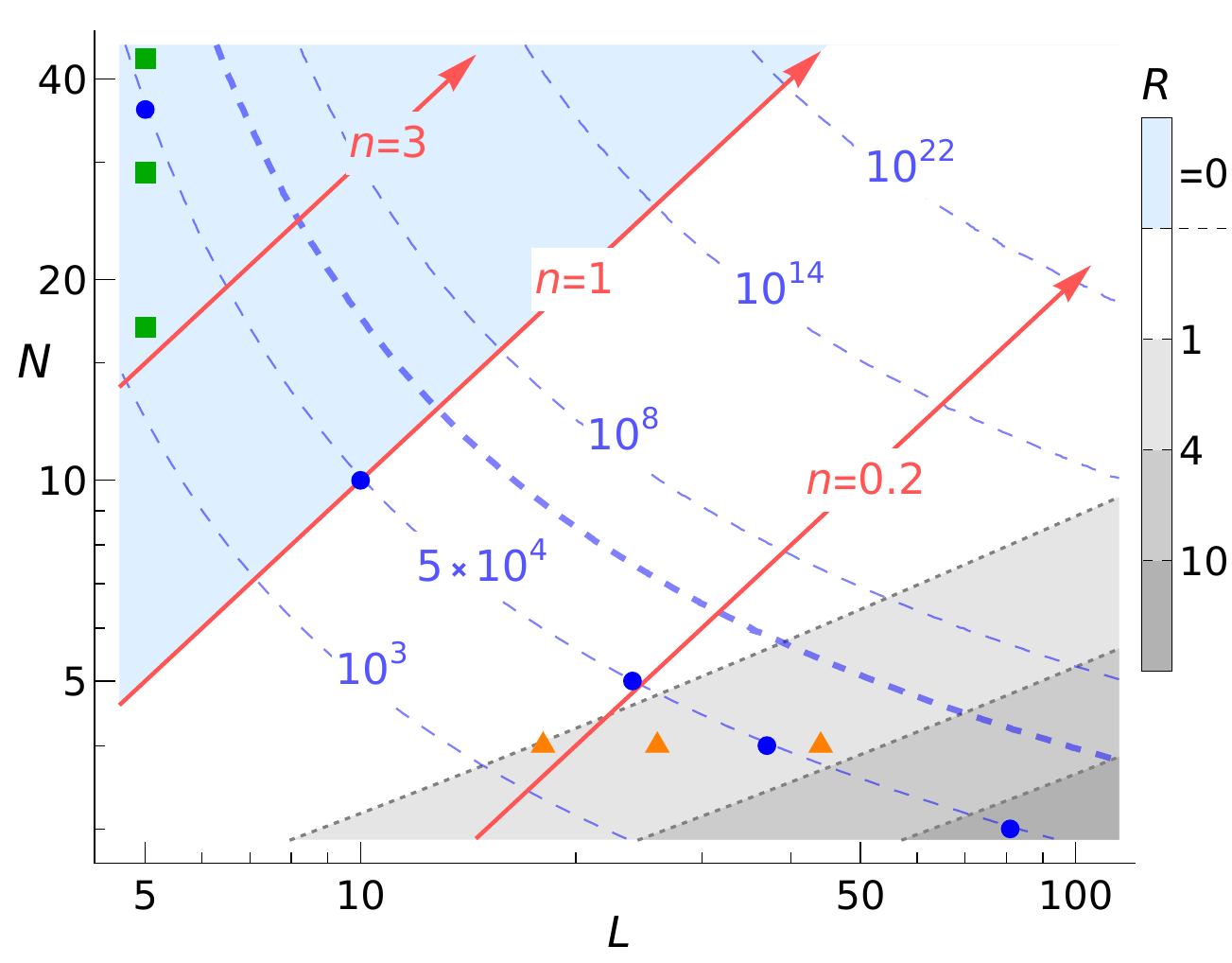}
 \caption{Size $\mathcal{N}$ [Eq.~\eqref{eq:dimH-}], indicated by labelled dashed contour lines, of the Hilbert space $\mathcal{H}^-$ as a function of $N$ and $L$ in log-log scale. (The thicker dashed contour line delimits the region accessed by our numerical simulations, $\mathcal{N}\lesssim 2\times10^6$.)
 Background colors identify areas within a range of values for the ratio $R$ [Eq.~\eqref{eq:defR}] of noninteracting to interacting basis elements. 
 Symbols mark the system configurations used in Figs.~\ref{fig:DPfixedHdim} (circles), \ref{fig:DPfixedNL}(a) (squares), and \ref{fig:DPfixedNL}(b) (triangles), lying along curves at constant $\mathcal{N}$, $L$ or $N$, respectively, while red straight lines highlight trajectories at fixed filling factor $n$.} 
 \label{fig:NLspace}
\end{figure}

The size of $\mathcal{H}^-$ is 
\begin{equation}
 \mathcal{N}=\frac{1}{2}\left[\begin{pmatrix} N+L-1 \\ N \end{pmatrix} -\Delta\right],
 \label{eq:dimH-}
\end{equation}
where the value of $\Delta$ depends specifically on the parity of $L$ and $N$ as given in Table \ref{tab:delta}.
\begin{table}
 \begin{tabular}{c|cc}
   & even $L$ & odd $L$ \\\hline 
   even $N$ & $\begin{pmatrix} (N+L-2)/2 \\ N/2 \end{pmatrix}$  & $\begin{pmatrix} (N+L-1)/2 \\ N/2 \end{pmatrix}$  \\
   odd $N$ &  0 & $\begin{pmatrix} (N+L-2)/2 \\ (N-1)/2 \end{pmatrix}$  \\
 \end{tabular}
 \caption{Values of $\Delta$ in Eq.~\eqref{eq:dimH-} in terms of the parity of $N$ and $L$.}
 \label{tab:delta}
\end{table}
The limit of infinite Hilbert space, $\mathcal{N}\to\infty$, can be reached by increasing $N$ or $L$ independently, or both simultaneously keeping the \emph{filling factor} (density)
\begin{equation}
  n=\frac{N}{L}
\end{equation}  
fixed, as illustrated in Fig.~\ref{fig:NLspace}. While the limit $L\to\infty$ at fixed $N$ should intuitively dilute the interaction term in the Hamiltonian, and hence lead to the absence of a chaotic regime, the latter is expected to persist along the path $N\to\infty$ at fixed $L$ as well as on trajectories at fixed $n$. 
Note, however, that the asymptotic growth of Hilbert space along these two paths is markedly different: While the route at constant filling factor witnesses an exponential increase of Fock configurations with particle number, $\mathcal{N}\sim [n^{-1}(1+n)^{1+1/n}]^N/\sqrt{N}$, trajectories at constant system size $L$ are characterized by the power-law dependence $\mathcal{N}\sim N^{L-1}$.

\begin{figure*}
 \includegraphics[width=.9\textwidth]{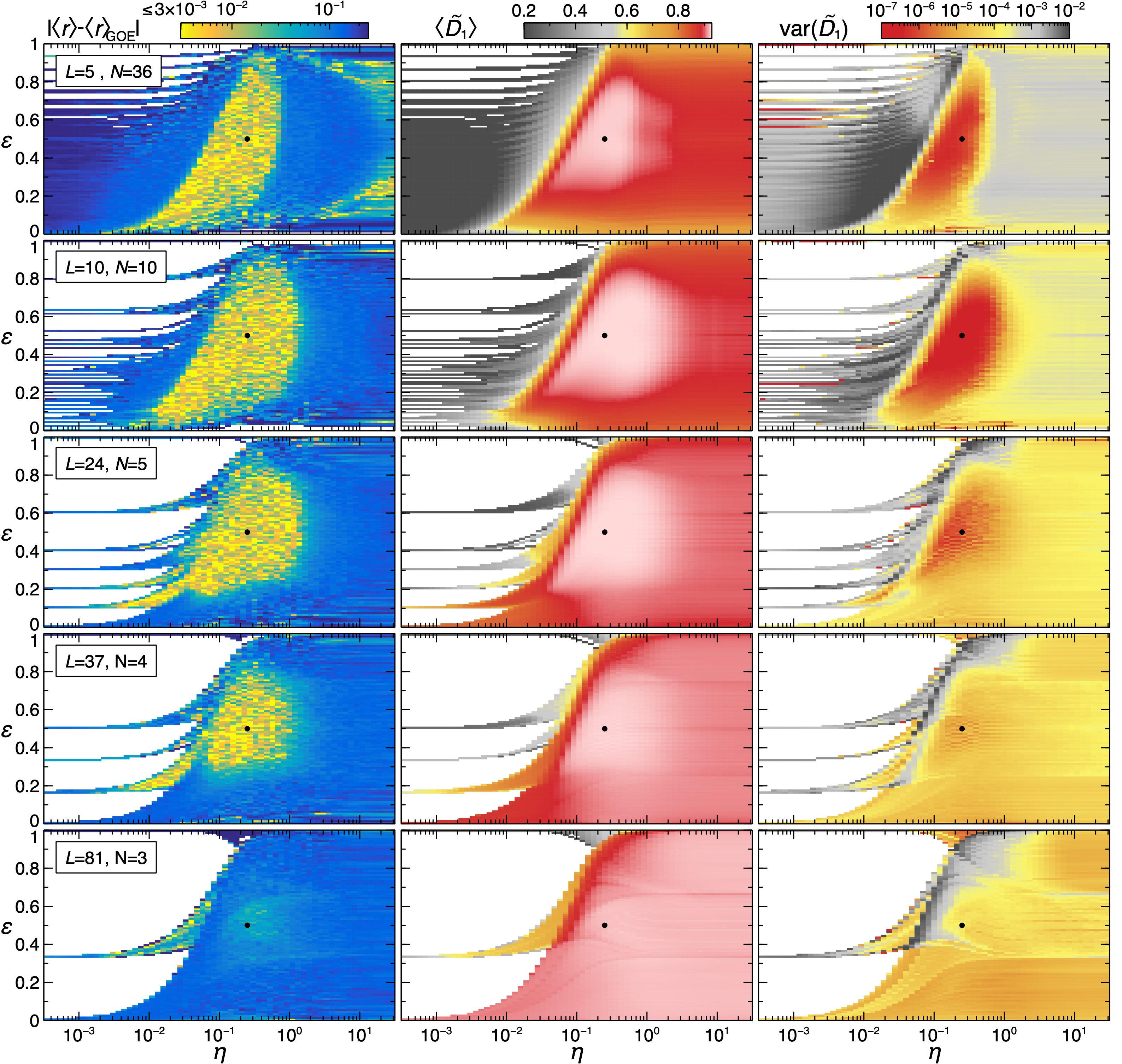}
\caption{Evolution of the chaotic phase for systems with comparable Hilbert space sizes ($\mathcal{N}\in[45\,600,49\,140]$) but different filling factor (from top to bottom, $n=7.2,1,0.21,0.11,0.04$) in terms of the mean level spacing ratio $\ar$ (left), mean fractal dimension $\aDq{1}$ (centre) and variance $\vDq{1}$ (right) in the interaction basis of $\mathcal{H}^-$, versus $\eta=J/UN$ and $\eps$.  The spectra were obtained for 64 equally spaced values of $\log\eta$, and divided into 100 bins of equal width along the $\eps$ axis. White areas highlight regions with an absence of energy levels. The black circle marks the point $\eps=0.5$, $\eta=0.25$, around which the data of Figs.~\ref{fig:varratios}, \ref{fig:D1diff}, and \ref{fig:PDFdiff}  are obtained.} 
 \label{fig:DPfixedHdim}
\end{figure*}

To investigate how quantum chaos in the BHH develops in $NL$-space, in Fig.~\ref{fig:DPfixedHdim}, we show an overview of the chaotic phase for five configurations with comparable Hilbert space sizes ($\mathcal{N}\simeq 5\times 10^4$, see the blue points in Fig.~\ref{fig:NLspace}) but with widely varying filling factors, $0.04\leqslant n\leqslant 7.2$. The figure shows the energy and $\eta$ resolved chaotic phase revealed by $\ar$, $\langle\Dq{1}\rangle$, and $\vDq{1}$.
We observe how the spectrally chaotic region, where $\ar\simeq\ar_\mrm{GOE}$, correlates with a marked increase of the fractal dimension, i.e., a delocalization tendency in Hilbert space, and with a very pronounced decrease of its fluctuation among close-in-energy eigenstates [$\vDq{1}$ drops by 4 orders of magnitude within the chaotic region for $n=1$].

In terms of $\ar$, the chaotic region seems to be most prominent around $n=1$ (second-row panels in Fig.~\ref{fig:DPfixedHdim}), whereas it undergoes a visible shrinking as the filling factor is further decreased at constant $\mathcal{N}$, and arguably a subtle narrowing also takes place for $n=7.2$ (first-row panels). This tendency is also visible in the evolution of $\vDq{1}$, which undergoes a very noticeable change. In fact, the disappearance of the chaotic phase for decreasing $n$ is earlier witnessed by the variance of the fractal dimension, and the vanishing of its sharp minimum, than by the average $r$ statistic (cf.~fourth-row panels in Fig.~\ref{fig:DPfixedHdim}). The dissolution of the `chaotic sea' is also evident from $\langle\Dq{1}\rangle$ (middle column in Fig.~\ref{fig:DPfixedHdim}), and as $n$ diminishes the values of the fractal dimension in the potentially chaotic region become progressively the same as those in the non-interacting limit ($\eta\to\infty$). In the chosen basis, the eigenstates in the non-interacting limit are greatly delocalized (although they are not extended ergodic as $\mathcal{N}\to\infty$), hence the homogeneously large values of $\langle\Dq{1}\rangle$ observed for $n=0.04$ (bottom row), despite the lack of chaotic fingerprints in $\ar$ and $\vDq{1}$. 

\begin{figure}
\includegraphics[width=\columnwidth]{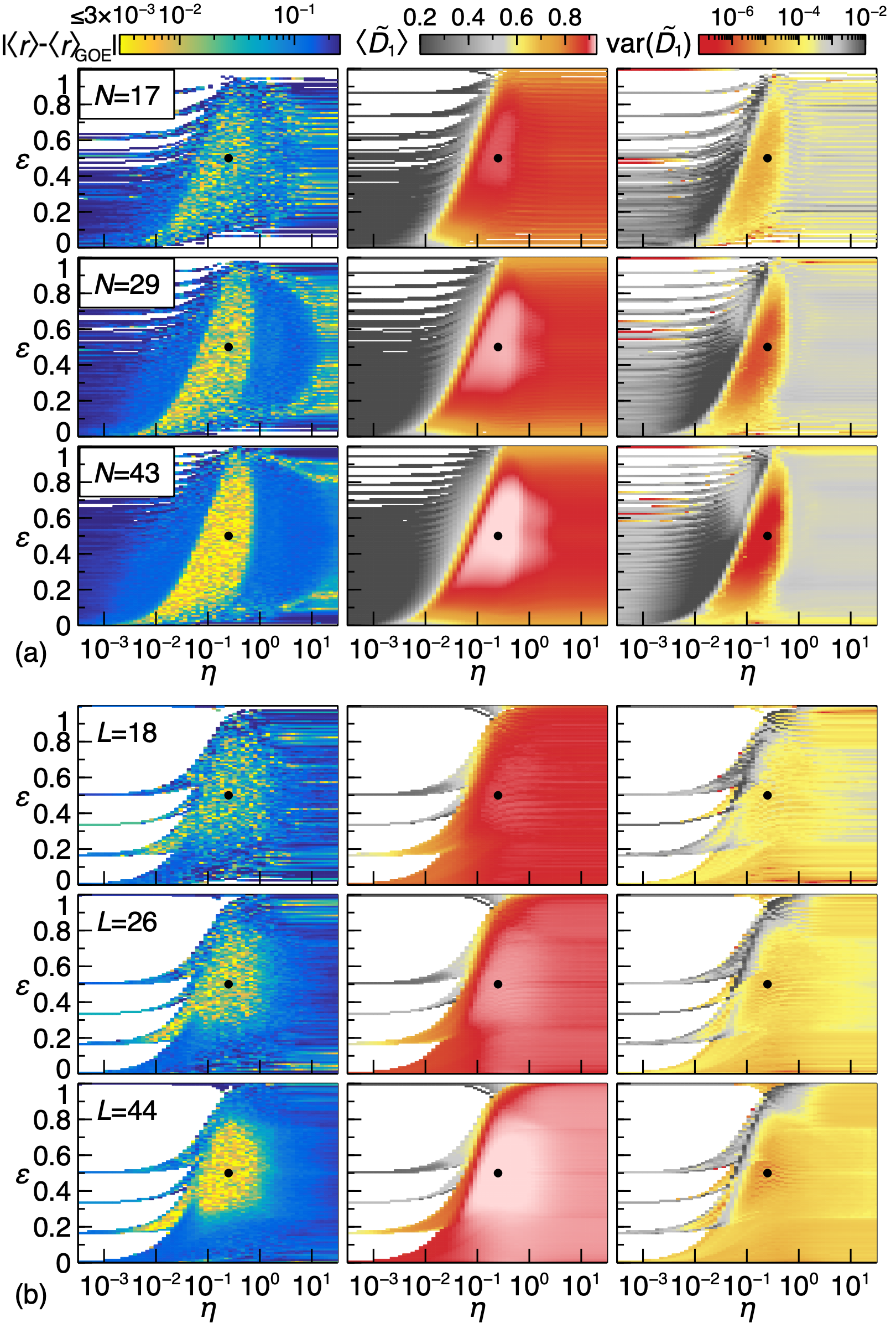}
\caption{Evolution of the chaotic phase for (a) fixed $L=5$ and $N=17,29,43$ ($\mathcal{N}=2970,20\,400, 89\,056$), and (b) for fixed $N=4$ and $L=18,26,44$ ($\mathcal{N}=2970, 11\,830, 89\,056$), using the same figures of merit as in Fig.~\ref{fig:DPfixedHdim}.} 
  \label{fig:DPfixedNL}
\end{figure}

These results indicate that for a given Hilbert space size, the most chaotic configuration occurs around unit filling factor. Let us check whether this picture holds as $\mathcal{N}$ is enlarged. In Refs.~\cite{Pausch2020,Pausch2021}, we demonstrated that the chaotic phase gets increasingly better defined (when comparing against GOE benchmarks) as the thermodynamic limit is approached along trajectories at constant $n=1$. A sample of the spectral and eigenvector features evolving along trajectories at constant $L$ or constant $N$ is shown in Fig.~\ref{fig:DPfixedNL} (see square and triangular symbols in Fig.~\ref{fig:NLspace}). 
As expected, along the trajectory at fixed $L$, the presence of a chaotic regime is enhanced for larger boson number (i.e., larger Hilbert space size), both in terms of the level statistics and in the drop of the variance of $\Dq{1}$ [Fig.\ref{fig:DPfixedNL}(a)]. However, along the trajectory at fixed $N$ [Fig.\ref{fig:DPfixedNL}(b)], no clear indication of the weakening of chaos is observed for the available system sizes: As $L$ is increased, the spectrally chaotic region gets better defined, and, although the presence of the chaotic phase in the eigenvector features is rather subtle, the $\aDq{1}$ value remains visibly higher than in the non-interacting limit, and the drop of $\vDq{1}$ slowly develops. Arguably, although one intuitively expects a complete absence of chaos for $L=\infty$, the evolution along this path is most likely non-monotonic: Chaos would first emerge as Hilbert space grows and the density of states becomes more continuous, giving rise to a denser spectrum bulk, to eventually fade as $L$ becomes large enough. It is nonetheless surprising that for the configuration $N=4$, $L=44$ ($\mathcal{N}=89\,056$), there is still no sign of such turning point, and although very slowly, the chaotic phase continues to get sharper. 
 
This slow behaviour can be understood by comparing the populations of non-interacting and interacting basis states in $\mathcal{H}^-$. 
The number of states of the form \eqref{eq:intbasis} with vanishing interaction energy is non-zero only for $L\geqslant N$. For simplicity, let us consider the case of even $L$ and odd $N$, noting that this assumption does not affect the conclusions obtained below. In this case, the number of non-interacting states reads $\begin{psmallmatrix} L \\ N \end{psmallmatrix}/2$, and the 
ratio of non-interacting to interacting configurations is 
\begin{equation}
 R=\left[2\mathcal{N}\begin{pmatrix} L \\ N \end{pmatrix}^{-1}-1 \right]^{-1}, \quad L\geqslant N, 
 \label{eq:defR}
\end{equation}
with $\mathcal{N}$ from Eq.~\eqref{eq:dimH-}. The evolution of $R$ as a function of $L$ and $N$ can be seen in Fig.~\ref{fig:NLspace}. An expansion for large $L$ reveals the following behaviour, 
\begin{equation}
 R= \frac{L}{N(N-1)}- 1/2 + O(L^{-1}),
 \label{eq:expR}
\end{equation}
indicating that, as expected, the ratio diverges as $L\to\infty$ for any fixed $N$ (i.e., the non-interacting configurations become dominant), but does so only linearly with the system size, and with a slope that decreases quadratically with the number of bosons. As can be seen in Fig.~\ref{fig:NLspace}, the configurations for $N=4$ considered in Fig.~\ref{fig:DPfixedNL}(b), still correspond to low values of the ratio, $1<R<4$. The slow linear growth of $R$ with $L$ implies that the disappearance of the chaotic phase along fixed-$N$ trajectories will not be observable for the numerically accessible system sizes.

From expansion \eqref{eq:expR}, one can approximate the functional form of $R$ isolines in $NL$-space, 
\begin{equation}
 N_R\simeq \frac{1}{2} + \frac{1}{2}\left(1  + \frac{8 L}{1 + 2R}\right)^{1/2},
\end{equation}
as indicated by dotted lines in Fig.~\ref{fig:NLspace}. The sublinear dependence of the latter boundaries on $L$ has an immediate important consequence: Any trajectory towards $\mathcal{N}\to\infty$ at constant non-vanishing filling factor $n$ ---no matter how arbitrarily small--- will reach $R=0$ and should exhibit the persistence of the chaotic phase. The asymptotic vanishing of $R$ at constant $n$ is exponential with $L$, 
\begin{equation}
 R\underset{L\to\infty}{\simeq} \sqrt{\frac{1+n}{1-n}} \left[(1+n)^{1+n}(1-n)^{1-n} \right]^{-L}, \quad n<1. 
\end{equation}

The influence of the manifold of non-interacting Fock configurations on the formation of the chaotic phase can be nicely observed in the $\ar$ density plots of Fig.~\ref{fig:DPfixedHdim}: As filling factor is reduced at constant Hilbert space size, the ratio $R$ increases and the manifold of non-interacting states, which is easily identified as the one emerging from zero energy at small $\eta$, gradually takes over the spectrally chaotic phase. 

\begin{figure*}
 \includegraphics[width=.32\textwidth]{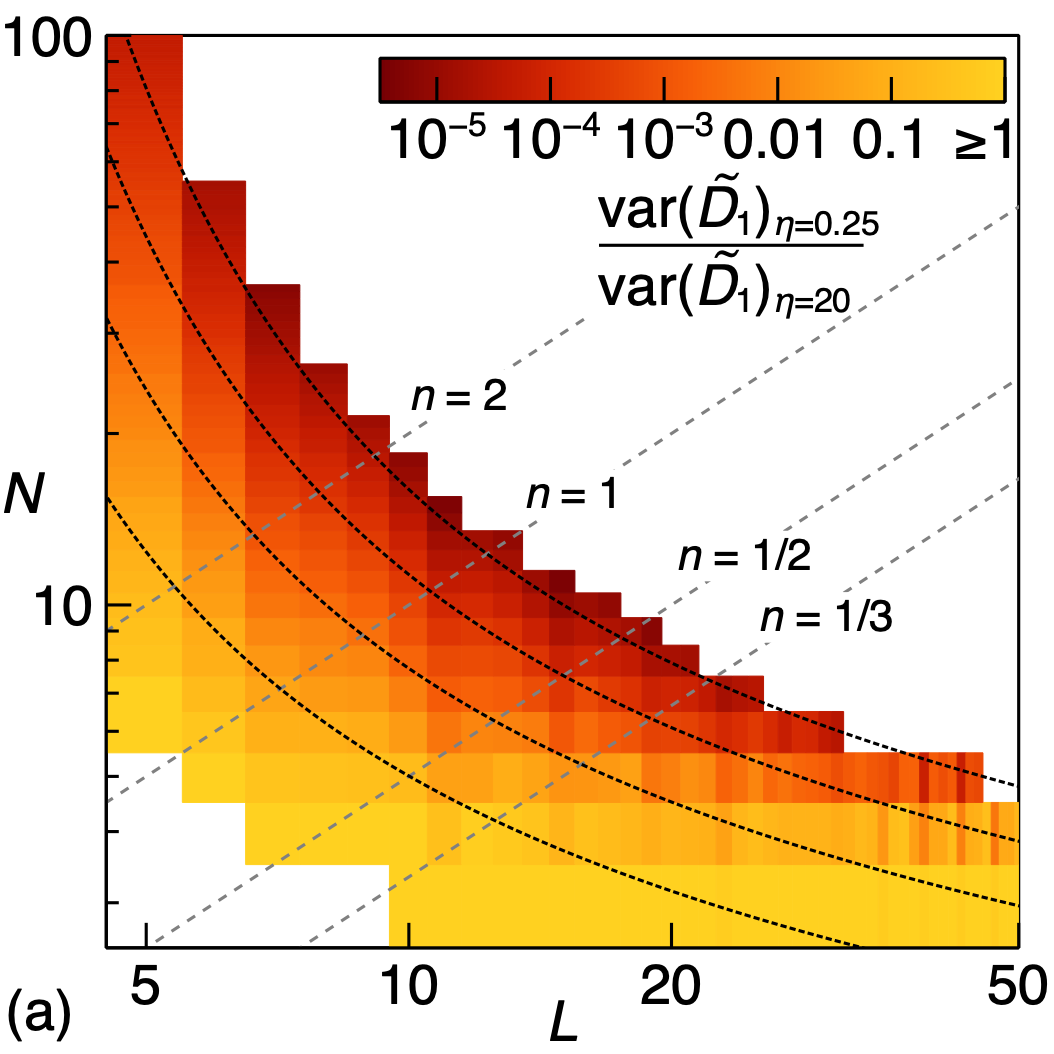}\hskip 2mm
 \includegraphics[width=.315\textwidth]{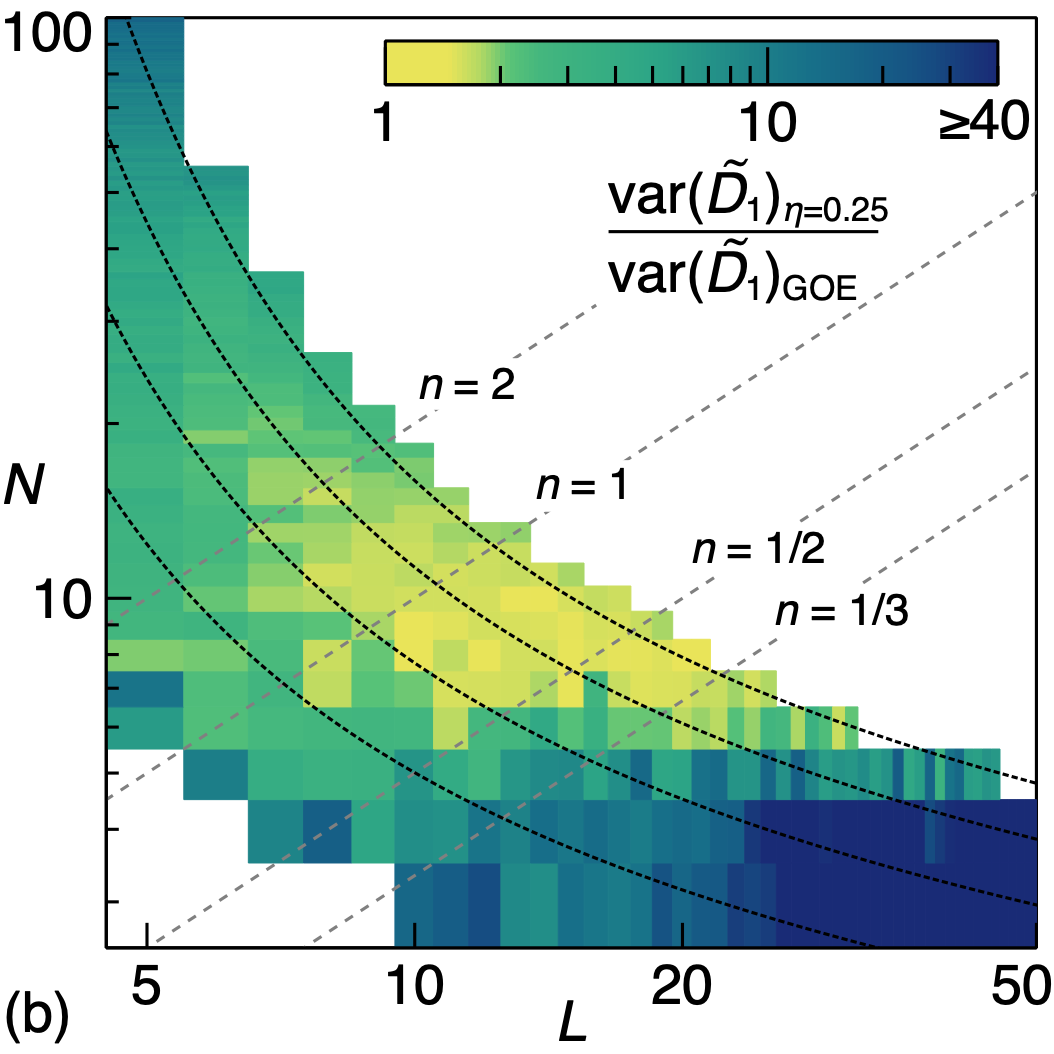}\hskip 2mm
 \includegraphics[width=.315\textwidth]{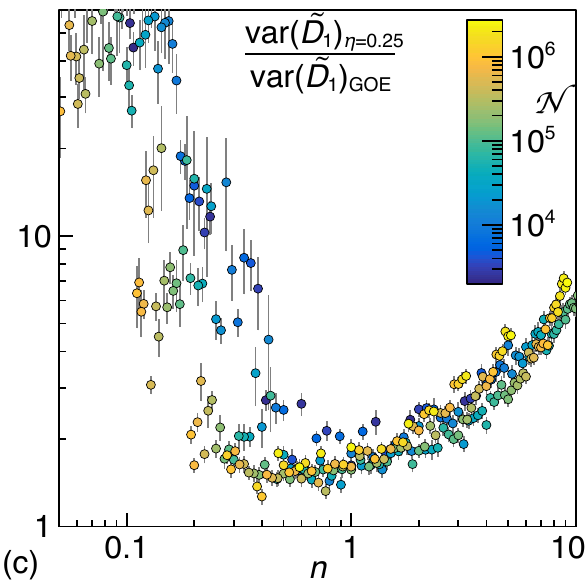}
 \caption{Development of the chaotic phase across $NL$-space characterized by the ratio of $\vDq{1}$ at the centre of the chaotic phase (100 eigenstates around $\eps=0.5$, averaged over $\eta\in[0.23,0.27]$) (a) to its value in the non-interacting limit, estimated from 100 eigenstates around $\eps=0.5$ at $\eta=20$, and (b) to the corresponding GOE value. Dotted curves mark the lines of constant Hilbert space size $\mathcal{N}=10^3,10^4,10^5,10^6$ (from bottom to top). 
 Panel (c) shows the ratio $\vDq{1}/\vDq{1}_\mrm{GOE}$ as a function of filling factor $n$ for different Hilbert space sizes. The estimation of $\vDq{1}$ at $\eta=20$ follows from a geometrical average over systems at fixed $L$, in order to smooth out $N$-parity related effects in the visualization.}
 \label{fig:varratios}
\end{figure*}

An overall perspective of the development of the chaotic region in $NL$-space from eigenvector features is provided in Fig.~\ref{fig:varratios}(a). There, the figure of merit is the contrast of the variance of $\Dq{1}$ at the centre of the chaotic phase, around 
$(\eta,\eps)=(0.25,0.5)$ 
(see black points in Figs.~\ref{fig:DPfixedHdim} and \ref{fig:DPfixedNL}), with respect to the large-$\eta$ region dominated by the many-particle non-interacting limit, $(\eta,\eps)=(20,0.5)$. The data clearly show the emergence of quantum chaos along trajectories at fixed $L$ and fixed $n$ as $\mathcal{N}\to\infty$. As discussed above, the slow disappearance of chaos at fixed $N$ is not visible for the accessible $\mathcal{N}$ except for the case $N=3$. 

Additionally, in Fig.~\ref{fig:varratios}(b), we compare $\vDq{1}$ to their corresponding GOE values (which we calculated analytically in Ref.~\cite{Pausch2020}), i.e., we quantify how chaotic the system is in relation to the RMT benchmarks. Interestingly, an optimal region, in the latter sense, appears between filling factors $n=1$ and $n=1/2$. The analysis of $\vDq{1}/\vDq{1}_\mrm{GOE}$ at fixed $\mathcal{N}$ versus filling factor in Fig.~\ref{fig:varratios}(c) furthermore demonstrates that this region extends from $n\simeq1$ towards lower filling factors the larger the Hilbert space (e.g., until $n\simeq 1/3$ for $\mathcal{N}\simeq 10^6$), whereas for $n>1$ the approach to GOE is not accelerated by increasing $\mathcal{N}$. These results indicate that the trajectories at fixed $n\lesssim 1$ are the fastest routes to quantum chaos in the BHH. 

\begin{figure}
\centering
\includegraphics[width=\columnwidth]{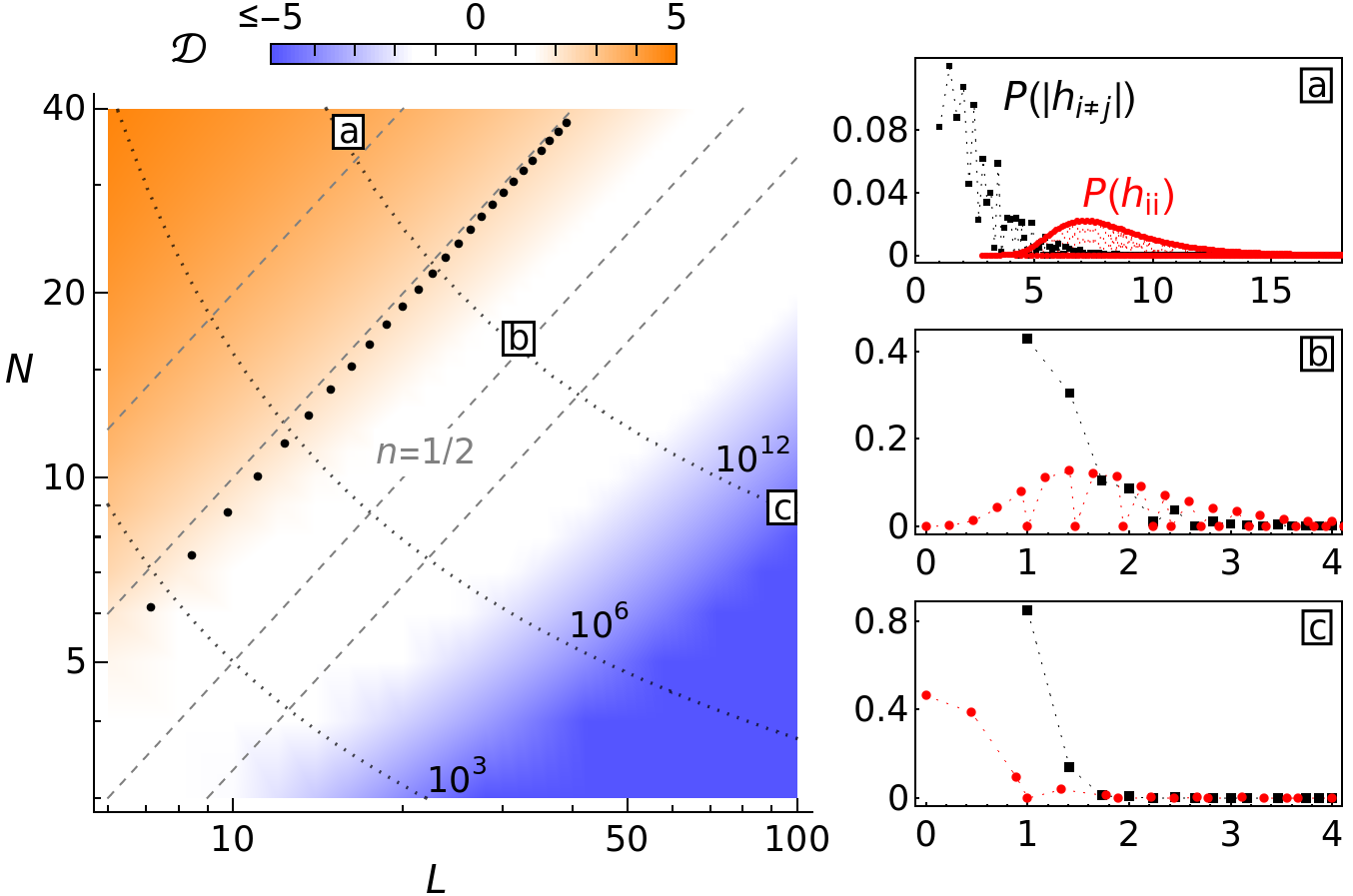}
\caption{Evolution in $NL$-space of the distance $\mathcal{D}$ [Eq.~\eqref{eq:Ddef}] between the off-diagonal and diagonal distributions of the BHH matrix elements at $\eta=0.25$. 
Dashed lines highlight trajectories at constant $n=2,1,1/2,1/3$ (from left to right), while dotted curves correspond to contours of constant Hilbert space size $\mathcal{N}$, and black points indicate the occurrence of maximum average Fock space connectivity $\mathcal{C}$ [Eq.~\eqref{eq:Cdef}] for fixed $\mathcal{N}$. Side panels show the corresponding distributions for the configurations marked by letters in the main plot. }
\label{fig:Hdist}
\end{figure}

The region of optimal approach to chaos mainly occurs within the range in $NL$-space characterized by small but non-vanishing values of the ratio $R$ [Eq.~\eqref{eq:defR}] of non-interacting to interacting Fock configurations in Hilbert space. A qualitative understanding of the location of this optimal chaotic region in $NL$-space may be obtained from the analysis of basic statistical features of the Hamiltonian matrix. First, we consider how the average connectivity in Fock space depends on $N$ and $L$, 
that under the simplifying assumption of even $L$ and odd $N$
can be seen to be (see Appendix~\ref{sec:connectivity})
\begin{equation}
	 \mathcal{C}=\frac{2N(L-1)}{N+L-1},
	 \label{eq:Cdef}
\end{equation}
and which will behave qualitatively the same for odd $L$ and/or even $N$.
The latter expression gives the average number of Fock states to which any basis state is connected by $H$. Intuitively, larger connectivity translates into stronger basis mixing, and hence, potentially, a faster development of quantum chaos. The connectivity is asymptotically bounded along trajectories at fixed $N$ or fixed $L$ by $2N$ or $2(L-1)$, respectively, while it remains unbounded along routes at constant filling factor. Despite the simple form of Eq.~\eqref{eq:Cdef}, its behaviour at constant Hilbert space size is not straightforward to read. A numerical analysis reveals that at fixed $\mathcal{N}$, the average Fock space connectivity is maximized at filling factors that asymptotically converge to $n=1$, as shown in Fig.~\ref{fig:Hdist}. 
Such maximal connectivity singles out the trajectory at $n=1$, and partially correlates with the location of the optimal chaos region, which however also extends to lower $n$, where the connectivity decreases.  

The analysis of the distributions of diagonal and non-vanishing off-diagonal $H$-matrix elements [with respect to the basis given in Eq.~\eqref{eq:intbasis}], $h_{ii}$ and $h_{i\neq j}$, respectively, for $\eta=0.25$ also offers some relevant information. Specifically, in Fig.~\ref{fig:Hdist}, we show 
\begin{equation}
 \mathcal{D}=\frac{\overline{h_{ii}}-\overline{|h_{i\neq j}|}}{\sigma(|h_{i\neq j}|)},
 \label{eq:Ddef}
\end{equation}
quantifying the distance between the mean values measured in terms of the standard deviation of the off-diagonal distribution. As can be observed, in terms of their average values, the distributions are closer  in a region centred around $n\simeq 1/2$. 
It is worth mentioning that $\mathcal{D}$ in the full Fock basis, i.e., without (anti)symmetrizing the states $\ket{n_1,\ldots,n_L}$, exhibits exactly the same behaviour.
Note that the variation of $\mathcal{D}$ is not just the result of a trivial shift of the whole diagonal distribution (which would have no consequences in the properties of the energies and eigenvectors), but rather reflects an overall change in the distributions: e.g., $P(h_{ii})$ broadens and develops a longer tail reaching larger values for high $n$, whereas the dominance of non-interacting configurations induces a drift of the most probable value towards zero and a faster decay for very small $n$. 
An example of the distributions for different configurations of comparable Hilbert space size is shown in the side panels of Fig.~\ref{fig:Hdist}.
We also checked that the analysis of the distribution widths in relation to the GOE definition [namely whether $\var(h_{ii})=2\var(|h_{i\neq j}|)$ holds] shows no distinctive behaviour in the optimally chaotic region.

Therefore, the region with densities around 
$1/2\lesssim n\lesssim 1$, 
where the fastest approach to quantum chaos is found, stands out according to the presented analysis based on Fock space connectivity and basic statistical properties of $H$. 

\section{Trajectories at constant $\boldsymbol{L}$ versus routes at fixed density}
\label{sec:pdfs}

\begin{figure}
\centering
\includegraphics[width=.9\columnwidth]{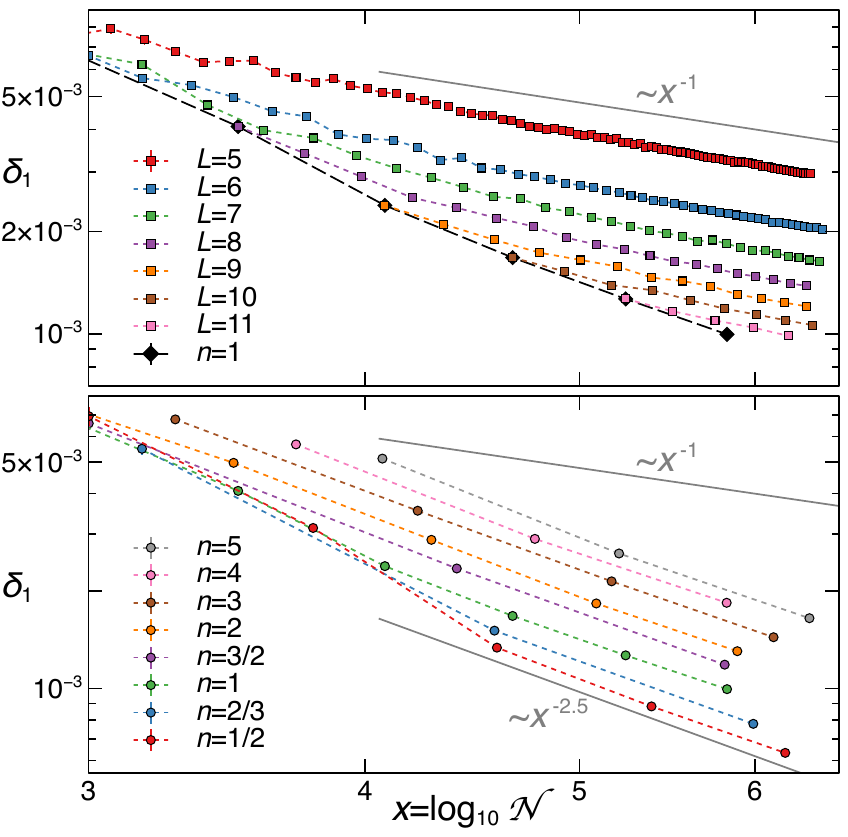}
\caption{Evolution of the distance $\delta_1(\mathcal{N})$ [Eq.~\eqref{eq:Delta1def}] versus $x\equiv \log_{10}\mathcal{N}$ in log-log scale, around $\eps=0.5$ and averaged over $\eta\in[0.23,0.27]$. The upper panel highlights the trajectories for fixed $L$ as the number of bosons and hence the Hilbert space size is increased (only for $N\geqslant L$ for clarity), while the lower panel indicates trajectories at different fixed filling factors. Solid gray lines mark the decays $x^{-1}$ and $x^{-2.5}$.}
\label{fig:D1diff}
\end{figure}

Figure \ref{fig:varratios}(b) exposes that the development of quantum chaos along trajectories at fixed $L$ is not as efficient as through paths at constant $n$: $\vDq{1}/\vDq{1}_\mrm{GOE}$ in fact exhibits an asymptotic increasing tendency at fixed system size [see the vertical trajectories at $L=5$ to $L=10$], revealing that the vanishing of $\vDq{1}$ in the thermodynamic limit follows a slower decay with $\mathcal{N}$ than the one for GOE. 
Such difference in the convergence towards the ergodic limit between the two sets of trajectories can also be observed in the average value of $\Dq{1}$. 
As we obtained in Ref.~\cite{Pausch2020}, GOE eigenvectors obey the asymptotic dependence 
\begin{equation}
 \aDq{1}_\mrm{GOE}= 1-\frac{c_1}{\ln\mathcal{N}}+O\left((\mathcal{N}\ln\mathcal{N})^{-1}\right), 
\end{equation}
with $c_1=2-\gamma-\ln 2\simeq 0.7296$, where $\gamma$ is Euler's constant. The data shown in Fig.~\ref{fig:D1diff} demonstrates that the difference
\begin{equation}
 \delta_1(\mathcal{N})=\aDq{1}_\mrm{GOE}-\aDq{1}
 \label{eq:Delta1def}
\end{equation}
along trajectories at constant $L$ vanishes asymptotically as $\delta_1\sim (\ln\mathcal{N})^{-1}$, indicating that for fixed system size $\aDq{1}$ indeed converges to $1$ and shares the analytical dependence of the dominant finite-size term with GOE, albeit with a manifestly different coefficient. On the other hand, at constant filling factor, the vanishing of $\delta_1$ with $\mathcal{N}$ is unmistakably faster (cf.~bottom panel in Fig.~\ref{fig:D1diff}), with no trace of the $(\ln\mathcal{N})^{-1}$ dependence, signalling that along fixed-$n$ trajectories $\aDq{1}$ bears the same dominant finite-size term as GOE, i.e., with the same coefficient.
Additionally, the asymptotic value of $\delta_1$ at a given $\mathcal{N}$ diminishes as the filling factor is lowered, reflecting a closer approach to GOE, in agreement with the observations for $\vDq{1}$ of Sec.~\ref{sec:chaosnu}.

\begin{figure}
	\centering
	\includegraphics[width=\linewidth]{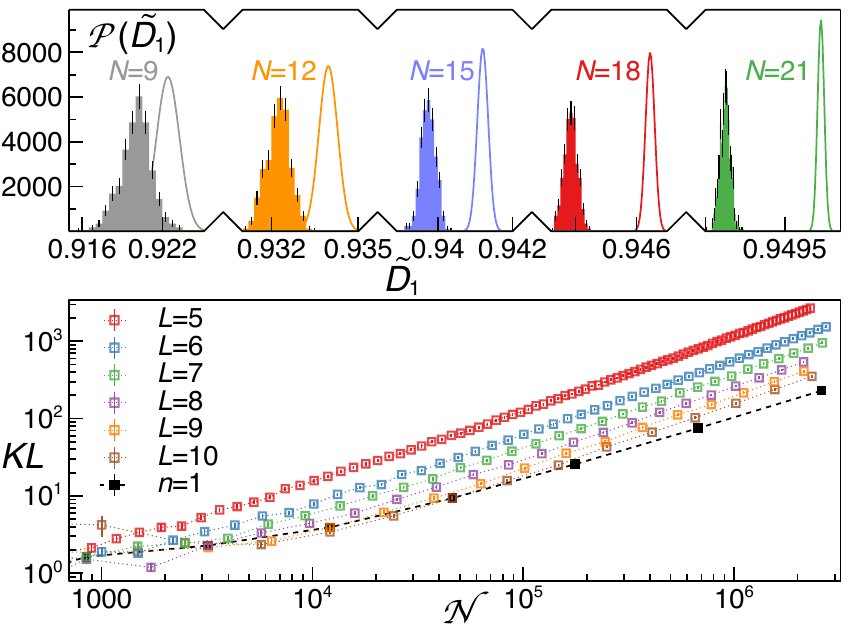}
	\caption{Evolution of the $\Dq{1}$ probability density function of BHH at fixed $L$ for increasing Hilbert space size $\mathcal{N}$ and comparison to GOE. The upper panel shows BHH histograms (filled) for $L=9$ and varying $N$, obtained from 500 eigenstates (100 states closest to $\eps=0.5$ at $\eta=0.23, 0.24, 0.25, 0.26, 0.27$), and the corresponding GOE distributions (solid lines) \cite{Pausch2020,Pausch2021,Pausch2022} (an $N$-dependent normalization factor is used to ease visualization). The lower panel displays the Kullback-Leibler divergence $KL(\P,\P_\textrm{GOE})$ 
	as a function of $\NN$ for constant $L\in[5,10]$ and for constant filling factor $n=1$.}
	\label{fig:PDFdiff}
\end{figure}

Despite this fast convergence to GOE observed for the mean and the variance of $\Dq{q}$ at constant density, we have shown in Refs.~\cite{Pausch2020,Pausch2021,Pausch2022}, for filling factor $n=1$, that, nevertheless, the full probability distributions of $\Dq{q}$ for BHH and GOE in fact become ever better distinguishable from one another when $\NN$ is increased. 
Given the enhanced distance of $\aDq{1}$ and $\vDq{1}$ from GOE along the trajectories at constant $L$, we expect this result to be even more pronounced when approaching $\NN\to\infty$ at fixed $L$.

Figure \ref{fig:PDFdiff}, upper panel, shows the evolution of the probability density functions of $\Dq{1}$ for fixed $L=9$ and increasing $N$ from $N=9$ ($\NN=12\,120$) to $N=21$ ($\NN=2.1\times10^6$), in comparison with the corresponding probability density functions for GOE. 
As we discussed in Refs.~\cite{Pausch2020,Pausch2022}, the GOE distribution is well approximated by a Gaussian. 
As the particle number and hence the size of Hilbert space increases, the BHH and GOE distributions 
depart from each other, 
eventually ending up clearly separated from one another for the largest $N$ considered.
This behaviour ensues since the distribution widths decrease with $\mathcal{N}$ faster than the mean values approach,
and is a reflection of model-dependent subleading finite-size corrections in $\aDq{1}$, as we discussed in Ref.~\cite{Pausch2021}.

To quantify the distance between the probability densities $\P(\Dq{1})$ and $\P_\textrm{GOE}(\Dq{1})$, 
we compute their Kullback-Leibler divergence, $KL(\P,\P_\textrm{GOE})$ [see Eq.~\eqref{eq:KL}]. 
This distance measure is shown, as a function of 
$\NN$, in the lower panel of Fig.~\ref{fig:PDFdiff}, for varying $N$ at constant $L\in[5,10]$, and for 
constant filling factor $n=1$. The increase of the Kullback-Leibler divergence with $\NN$ confirms 
the departure of the probability distributions shown in the upper panel and reveals furthermore that the distributions are further away 
from each other the smaller $L$ is, in agreement with the corresponding larger values 
of $\delta_1$ observed in Fig.~\ref{fig:D1diff}.
This result can be readily explained from the connectivity of the underlying Fock basis states, which 
is mediated by the tunneling Hamiltonian and 
yields maximally $2(L-1)$ transitions from any given Fock state [compare Eq.~\eqref{eq:Cdef}]. 
In GOE, on the other hand, transitions are allowed between any pair of states. 
Hence, at constant $\NN$ and $n>1$, the smaller $L$ is, 
the further the Fock space structure imposed by BHH deviates from that of GOE, and consequently the eigenstates would be expected to bear stronger signatures of non-universal features. 
At constant density, however, connectivity grows with $\NN$, and, consequently, the departure between BHH and GOE distributions is slower than at fixed $L$. 

\section{Conclusions}
\label{sec:conclusions}
We have studied the dependence of the chaotic phase of the Bose-Hubbard Hamiltonian (BHH) on particle number $N$ and system size $L$, and its evolution as the limit of infinite Hilbert space is approached along different directions. The combined analysis of spectral statistics and eigenvector structure, in terms of the generalized fractal dimension $\Dq{1}$, confirms that the chaotic phase develops as $N$ increases, either at fixed system size or at constant density ---no matter how arbitrarily small---, albeit in a quantifiable different way. The observed slow disappearance of chaos along trajectories at constant boson number is explained by the corresponding slow linear divergence of the ratio of non-interacting to interacting basis configurations on these paths. 

The comparison of the fractal dimension's fluctuation for near-in-energy eigenstates against the GOE values unveils the existence of an optimal region for the emergence of quantum chaos in $NL$-space for densities $1/2\lesssim n \lesssim 1$ [Fig.~\ref{fig:varratios}(b)], and shows that the trajectories at such constant $n$ are the fastest routes to chaos in the BHH. The demonstrated qualitatively basis independent behaviour of $\vDq{q}$ \cite{Pausch2020,Pausch2021,Pausch2022} makes this finding a fundamental property of the BHH. 

The path $N\to\infty$ at fixed $L$ leads to a slower convergence of the chaotic phase towards random matrix theory benchmarks. Despite this convergence, in terms of the distribution of the eigenstate generalized fractal dimensions, the ergodic phase of the BHH becomes more distinguishable from random matrix theory for larger Hilbert space. Such departure from GOE is actually faster than along trajectories at constant filling factor. 

The optimal chaotic region in $NL$-space exhibits distinct features in terms of Fock space connectivity and basic statistical properties of the Hamiltonian matrix. 
While further ingredients, such as correlations among the BHH matrix elements, may play an important role in the formation of the chaotic phase, 
these results provide an elementary foundation on which a deeper understanding of this optimal behaviour may be built. 

\begin{acknowledgments}
A.R. thanks M.~Rigol for helpful discussions.
The authors acknowledge support by the state of Baden-W\"urttemberg through bwHPC and the German Research Foundation (DFG) through Grants No.~INST 40/467-1 FUGG (JUSTUS cluster), No.~INST 40/575-1 FUGG (JUSTUS 2 cluster), and No.~402552777.
E.G.C.~acknowledges support from the Georg H.~Endress foundation.
A.R. and L.P.~acknowledge support by Spanish MCIN/AEI/10.13039/501100011033 through Grant No.~PID2020-114830GB-I00. 
\end{acknowledgments}

\appendix
\section{Average connectivity $\mathcal{C}$}
\label{sec:connectivity}
Here, we provide the derivation of Eq.~\eqref{eq:Cdef}.
We consider the subspace $\mathcal{H}^{-}$ with odd $N$ and even $L$, for simplicity. 
In this case, each pair made up of one Fock state $\ket{n_1,\ldots,n_L}$ and its reversed $\ket{n_L,\ldots,n_1}$ corresponds uniquely to a basis state $\ket{\vec{n}}$ of $\mathcal{H}^{-}$, as defined in Eq.~\eqref{eq:intbasis}, and hence the connectivity of $\ket{\vec{n}}$ is the same as that of $\ket{n_1,\ldots,n_L}$. Furthermore, since the size of $\mathcal{H}^{-}$ is exactly half of the total Hilbert space, the average connectivities in both spaces coincide. Recall that the size of full Hilbert space reads 
\begin{equation}
 \mathcal{S}(L,N)=\binom{N+L-1}{N}.
\end{equation}

Let us consider first the simpler case of periodic boundary conditions. Any Fock state $\ket{n_1,\ldots,n_L}$ connects to two other states for each site with a non-vanishing population. Since all sites are equivalent, one can simply evaluate the connectivity at any site and then multiply by the number of sites. The number of states with a given occupation $k$ at one site $j$ is given by the size of Hilbert space for a system with $L-1$ sites and $N-k$ bosons. Summing over $1\leq k\leq N$ yields the number of all states where site $j$ has a nonvanishing connectivity.
Therefore, 
\begin{equation}
 \mathcal{C}_\text{PBC}=\frac{2L}{\mathcal{S}(L,N)}\sum_{k=1}^N \mathcal{S}(L-1,N-k)=\frac{2NL}{L+N-1}.
\end{equation}

In the case of hard-wall boundary conditions, the only difference is that populated edge sites only contribute with a unit factor to the connectivity, i.e., instead of $2L$ one has an overall multiplicative factor of $2(L-1)$, yielding 
\begin{equation}
 \mathcal{C}=\frac{2N(L-1)}{L+N-1}.
\end{equation}

%

\end{document}